\documentclass[12pt]{iopart}
\usepackage{graphicx}
\begin{document}

\title{Photons from nuclear collisions at RHIC energies}

\author{Charles Gale$^{1}$, Simon Turbide$^{1}$\footnote{Current address: Defense Research and Development Canada, 2459 Pie-XI Nord, Val-B\'elair, QC, Canada G3J 1X5}, Evan Frodermann$^{2}$ and Ulrich Heinz$^{2}$}

\address{$^{1}$ Department of Physics, McGill University, 3600 University Street, Montreal, QC, Canada H3A 2T8}
\address{$^{2}$ Department of Physics, The Ohio State University, Columbus, OH 43210, USA}

\ead{gale@physics.mcgill.ca}
\begin{abstract}
We model the hot and dense strongly interacting mater produced in high energy heavy ion collisions using relativistic hydrodynamics. Several different sources of real photons produced during these collisions are considered and their relative importance is assessed. We include contributions from QCD jets, which are allowed to loose and gain energy as they proceed through the hot matter. This is treated within the AMY formalism. We obtain photon spectra, $R_{AA}^\gamma$, and $v_2^\gamma$ in agreement with measurements performed by the PHENIX collaboration.

\end{abstract}


\section{Introduction}
The measurement of electromagnetic observables in relativistic nuclear collisions has constituted a vigorous experimental effort at different facilities around the world for several decades, and it is fair to write that this program has lived up to its promises. Indeed, photons (real and virtual) carry information complementary to that in hadronic observables, as they suffer essentially no final state interaction. One may show \cite{KG} that the photon emission rate in a finite-temperature medium is directly related to the in-medium retarded self-energy of the photon: a valuable quantity. However, the measured signal is comprised of several contributions which have to be under simultaneous theoretical control before significant claims can be substantiated.

\section{Photon production}
\subsection{Non-thermal processes}
As in nucleon-nucleon interactions, nucleus-nucleus collisions will produce prompt photons during the early collisions of the nuclear constituents. The prompt (direct + fragmentation) photon spectrum can be calculated in $p p$ collisions at next-to-leading order \cite{aur}. The direct contribution comprises both Compton and annihilation components:
\begin{eqnarray}
E \frac{d \sigma_{\rm prompt}^{pp}}{d^3 p} = \sum_{a, b, c} \int dx_{a} d x_{b} f_{a} (x_{a}, Q) f_{b} (x_{b}, Q) K_{\rm dir} \, E \frac{d \sigma_{a c \to c \gamma^{*}}}{d^{3}p} (p, Q)\nonumber \\
+ \sum_{a, b, d} \int dx_{a} d x_{b} f_{a} (x_{a}, Q) f_{b} (x_{b}, Q) \int \frac{d z}{z^{2}} K_{\rm frag} \frac{d \sigma_{ab \to c d}}{d^{2} q_{T} dy} (Q) D_{\gamma^{*}/c} (z, Q_{F})
\end{eqnarray}
The NLO effects are contained in the K-factors;  their list of arguments has been omitted for brevity. In principle, the prompt photon spectrum depends on the choice of renormalization, factorization ($Q$), and fragmentation ($Q_{F}$) scales; in the above the first has been set equal to the second. Choosing them to be $ p_{T}/\sqrt{2}$, where $p_{T}$ is the photon transverse momentum, yields a photon spectrum  in proton-proton collisions in good agreement with observation \cite{TGFH}. This can be then be used as a baseline for calculations of photon production in nucleus-nucleus collisions. In this case, isospin effects are included in the parton distribution function, by using the parton species for the proton or for the neutron, as appropriate. Shadowing corrections are also taken into account.

\subsection{Thermal processes}
In a thermal environment, whether composed of partons or of hadrons in the confined phase, the photons produced are calculated simply by performing an integral over the space-time history of the strongly interacting system:
\begin{eqnarray}
E \frac{d^{3} N_{\rm thermal}^{\gamma}}{d^{3} p} = \int d^{4}x\, E_{0} \frac{d^{3} R_{\rm thermal}^{\gamma}}{d^{3} p_{0}}
\end{eqnarray}
The differential photon production rate, $R$, is evaluated in the local thermal rest frame, and at a given time will consist of partonic and/or hadronic contributions.

Another process which depends on the temperature is that of jet-thermal photon production. This happens when a jet crossing the hot medium undergoes a photon-producing interaction with a thermal parton.  This can further be divided in two classes, those which will benefit from collinear enhancements, and those which will not \cite{TGFH}. All of the rates introduced so far need to be integrated using with  a realistic time-evolution model of the relativistic nuclear collision. We will use  a two-dimensional relativistic hydrodynamical approach - AZHYDRO - which provides a good global description of hadronic observables \cite{KH} and, importantly, is not tuned to specifically reproduce electromagnetic measurements. 

\section{Jet energy loss and gain}
The initial jet energy profiles calculated from QCD are evolved in time according to AMY \cite{amy}. The evolution of the parton energy profile, $P(E, t)$, is determined by solving a set of coupled Fokker-Planck rate equations which have the following generic form:
\begin{eqnarray}
\frac{d P(E)}{dt} = \int d\omega \left[ P(E+\omega) \frac{d\Gamma (E+ \omega, \omega)}{d\omega dt} - P(E) \frac{d\Gamma (E, \omega)}{d\omega dt}\right]
\end{eqnarray}
The transition rate for processes where partons of energy $E$ loose energy $\omega$ is $d\Gamma (E, \omega)/d\omega dt$. The energy gain channels are included through the possibility for $\omega$ to be negative. Also, AMY smoothly handles the crossover between the Bethe-Heitler and Landau-Pomeranchuk-Migdal regimes \cite{TGJM}. A free parameter of the model is $\alpha_s$, the strong coupling,  which is adjusted to reproduce $R_{AA}^{\pi^0}$. 

\section{Results and conclusions}
\begin{figure}
\begin{center}
\includegraphics[width=5.9cm,angle=-90]{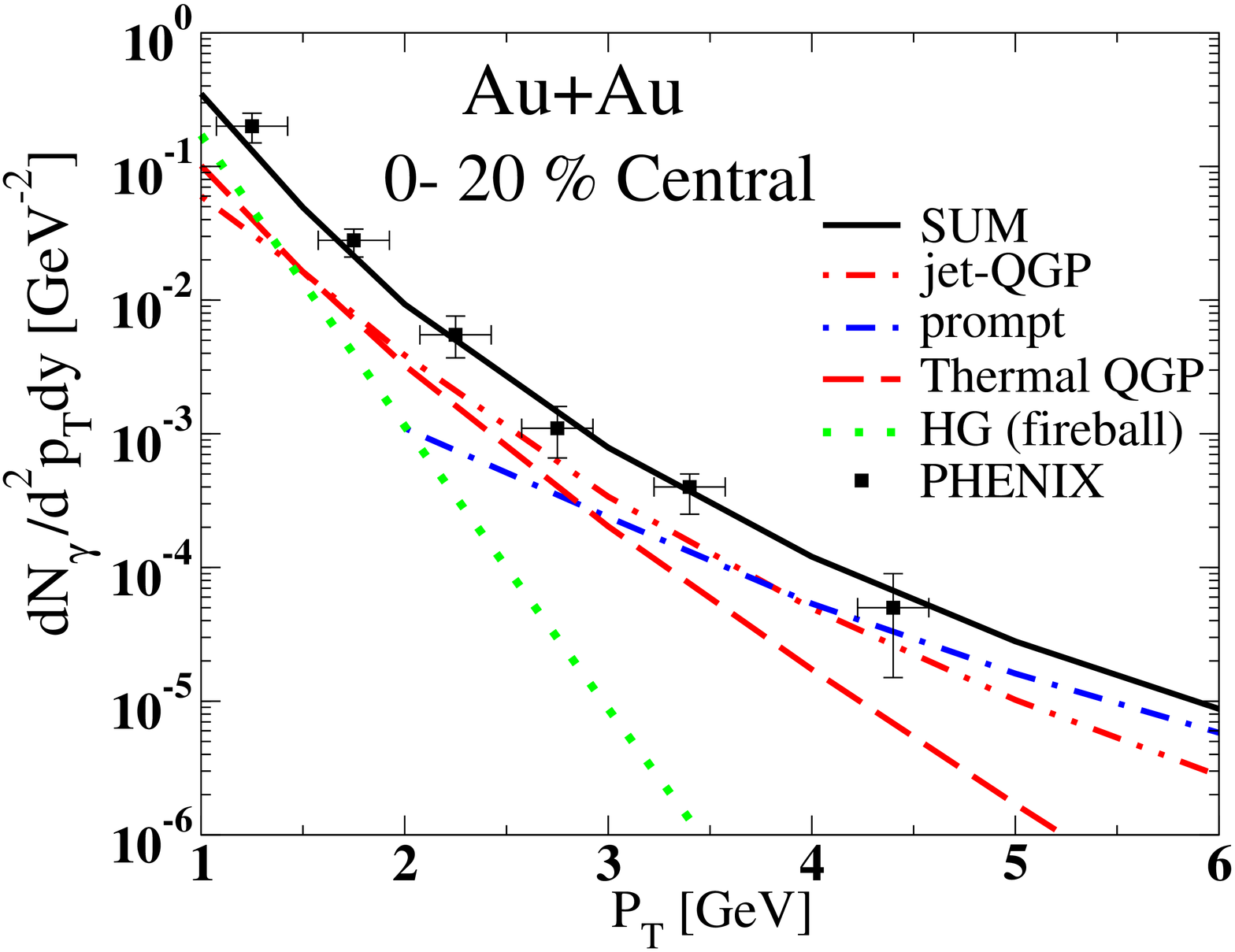}
\includegraphics[width=5.9cm,angle=-90]{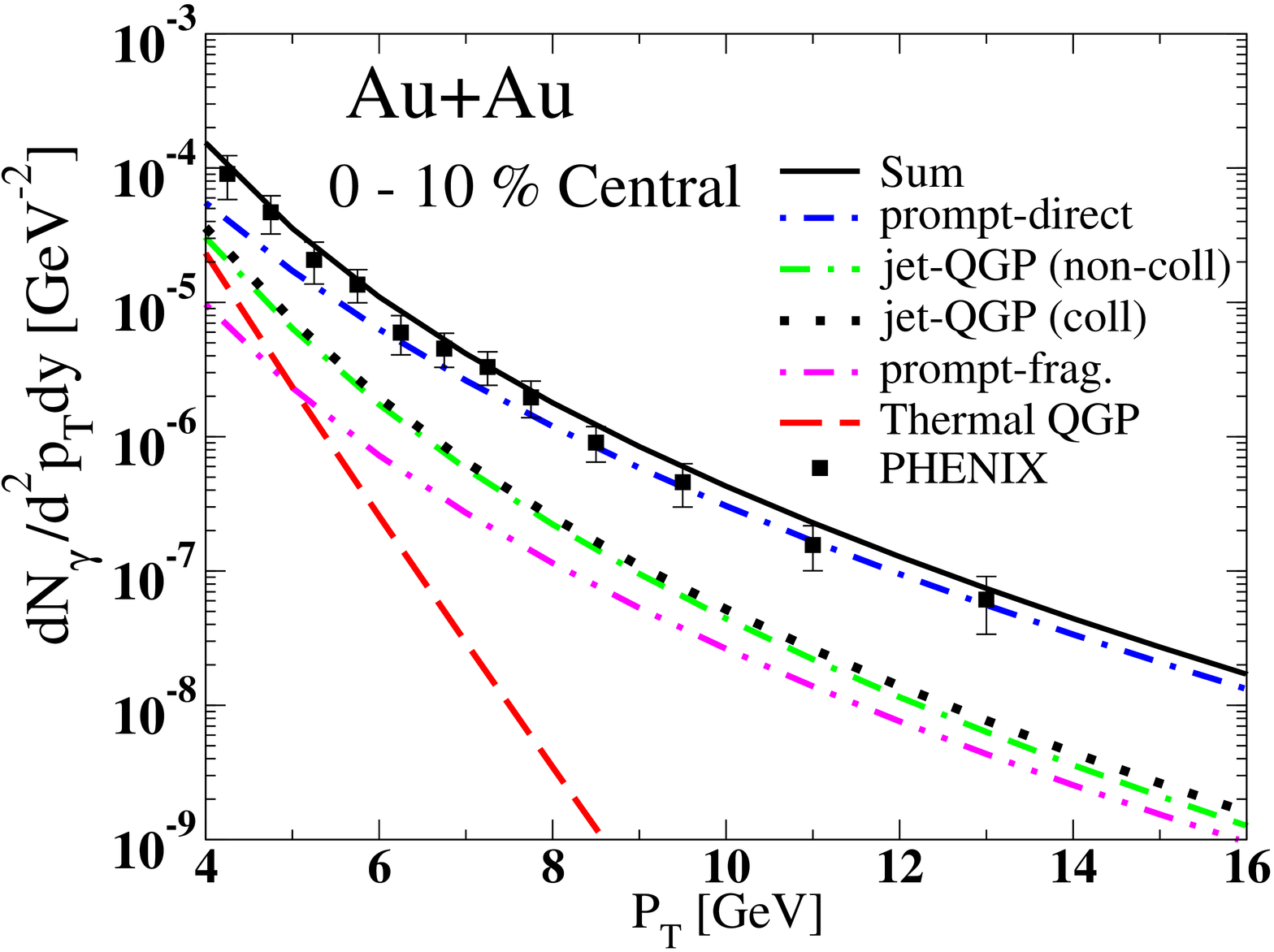}
\caption{Spectra of produced photons at RHIC, for two different centrality classes. The different elements of the theoretical calculations are shown separately. Data are from the PHENIX collaboration \cite{phe1,phe2}.}
\label{pho_spec}
\end{center}
\end{figure}

\begin{figure}[!h]
\begin{center}
\mbox{\includegraphics[width=6cm,angle=-90]{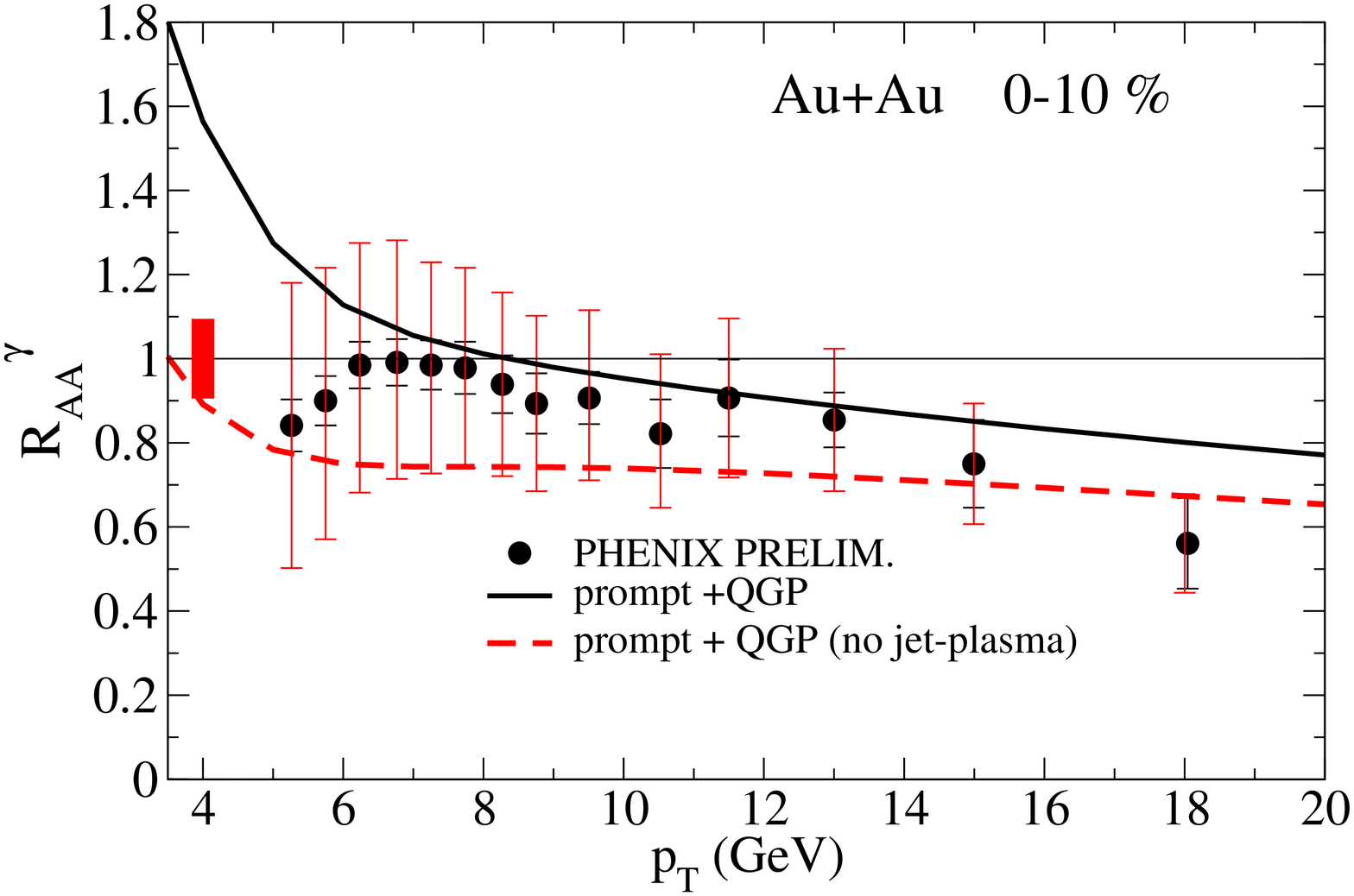}
\vspace*{2cm}\includegraphics[width=6.5cm,angle=-90]{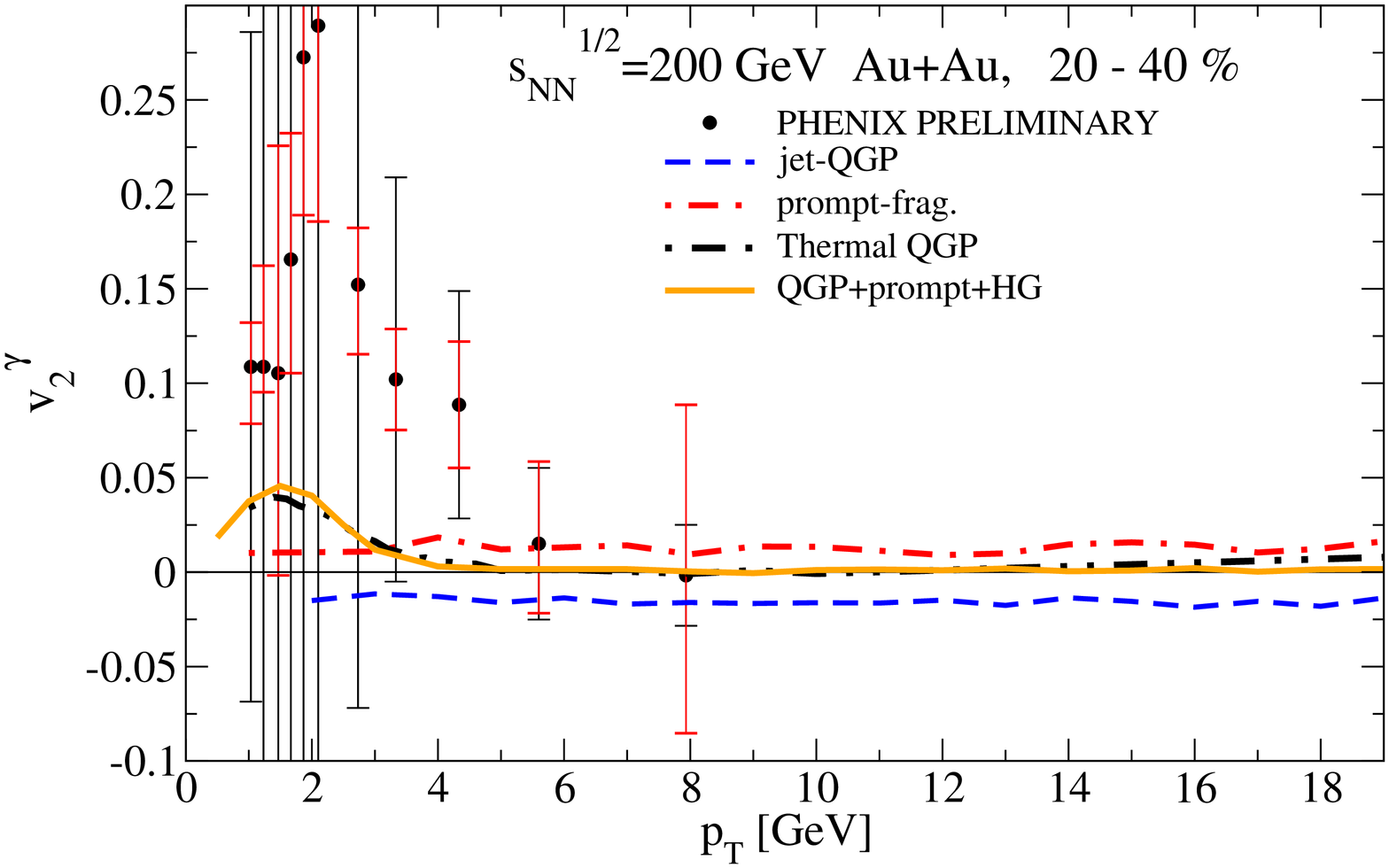}}
\caption{The nuclear modification factor of produced photons at RHIC (left panel), and the photon azimuthal anisotropy (right panel), for two different centrality classes. For $R_{AA}^\gamma$, cases with and without jet-plasma photons are shown. The different components of the photon contribution and their azimuthal anisotropy are shown on the right. Data are from the PHENIX collaboration.}
\label{R_AAv2}
\end{center}
\end{figure}

The calculated photon spectra, together with their different components, are shown in Figure \ref{pho_spec}. The reader may judge of the good agreement between theory and data. No parameters have been tuned to generate these figures. 
Two other useful representations of the experimental data are displayed in Figure \ref{R_AAv2}. The first (left panel) is the ubiquitous $R_{AA}^\gamma$ variable, while the second is a plot of the photon azimuthal asymmetry, $v_2^\gamma$.
In both panels, the data does not yet permit an accurate determination of model parameters and ingredients. However, the fact that $R_{AA}^\gamma <$ 1 at high values of $p_T$ is a consequence of the fragmentation photons being affected by the energy loss of the fragmenting jets (as well as by an isospin effect \cite{TGFH}): should this trend be confirmed experimentally, a link would exist between the nuclear modification factor of photons and that of strongly interacting particles also borne out of fragmentation. 

Turning to $v_2^\gamma$, one observes that the azimuthal anisotropy associated with the jet-plasma photons is indeed negative, as originally suggested \cite{TGF}. However, the net value of $v_2^\gamma$ appears to be essentially zero for $p_T >$ 4 GeV. The reason for this is two-fold. First, the geometric anisotropy of the medium in the realistic 2D+1 model is rather small initially, and then proceeds to shrink as time evolves. Second, the geometry of the initial jet profile is evaluated here with a realistic Woods-Saxon distribution, rather than using a hard sphere profile. However, these results do show that the photon $v_2$ is rather sensitive to early time dynamics in relativistic nuclear collisions: precise measurements thus have the potential to stringently constrain the time-evolution scenarios. 

A global picture is emerging where the physics of  QCD jets and the generation of electromagnetic radiation in nuclear collisions  need to be treated simultaneously and consistently. 
The results displayed here are being generalized to include energy loss by elastic collision processes \cite{qin1}, and the effects of a 3D+1 hydrodynamical evolution.

\end{document}